\documentclass[12pt]{iopart}
\pdfoutput=1
\usepackage{amssymb}
\usepackage{flushend}
\usepackage{cite}
\usepackage{graphicx}
\usepackage{iopams}  
\usepackage{algpseudocode} 
\usepackage{url}

\def \({\left(}
\def \){\right)}
\def \[{\left[}
\def \]{\right]}

\newcommand{\cv}[1]{\underline{#1} \,}
\newcommand{\bF}{{\textbf {F}}}
\newcommand{\bR}{{\textbf {R}}}

\newcommand{\bx}{{\textbf {x}}}

\newcommand{\by}{{\textbf {y}}}
\newcommand{\bv}{{\textbf {v}}}

\newcommand{\bxigma}{{\boldsymbol{\Sigma}}}
\newcommand{\bxii}{{\boldsymbol{\xi}}}

\begin{document}

\title[AMP with structured operators]{Approximate message-passing with
  spatially coupled structured operators, with applications to
  compressed sensing and sparse superposition codes}

%\author{Jean Barbier$^\dag$, Florent Krzakala$^{\dag,\star,\ddag}$ and Christophe Sch\"ulke$^{\ddag,\cup,\cap}$}
\author{Jean Barbier$^\dag$, Christophe Sch\"ulke$^{\cup,\cap}$ and Florent Krzakala$^{\dag,\star}$}

\address{$\dag$ Laboratoire de Physique Statistique, CNRS UMR 8550,
  Ecole Normale Sup\'erieure \& Universit\'e Pierre et Marie Curie, 24
  rue Lhomond, 75231 Paris, France} 
\address{$\star$ Sorbonne Universit\'es, UPMC Universit\'e Paris 06, UMR 8550 75231, Paris, France} 
%\address{$\ddag$ Laboratoire de  Physico-Chimie Th\'eorique, 10 rue Vauquelin, 75231 Paris, France}
\address{$\cup$ Universit\'e Paris Diderot, Sorbonne Paris Cit\'e, 75205 Paris, France}
\address{$\cap$ Dipartimento di Fisica,
  Sapienza Universit\`a di Roma, p.le A. Moro 2, 00185 Rome, Italy}

\ead{jean.barbier@ens.fr, christophe.schulke@espci.fr, florent.krzakala@ens.fr}

\begin{abstract}
We study the behavior of Approximate Message-Passing, a solver for
  linear sparse estimation problems such as compressed sensing, when
  the i.i.d matrices ---for which it has been specifically designed---
  are replaced by structured operators, such as Fourier and Hadamard
  ones. We show empirically that after proper randomization, the
  structure of the operators does not significantly affect the
  performances of the solver. Furthermore, for some specially designed
  spatially coupled operators, this allows a computationally fast and
  memory efficient reconstruction in compressed sensing up to the
  information-theoretical limit. We also show how this approach can be
  applied to sparse superposition codes, allowing the Approximate
  Message-Passing decoder to perform at large rates for moderate
  block length.
\end{abstract}

\newpage

Sparse reconstruction problems have recently witnessed a burst of activity
 in a wide range of applications spanning from signal processing
\cite{Donoho:06} to coding theory \cite{barron2010sparse}. In
particular, compressed sensing (CS) techniques
\cite{CandesRombergTao06, Donoho:06} have suggested entirely new ways
of capturing signals and images, such as in the single-pixel camera
\cite{duarte2008single}. 

The physics community became interested in these topics due to the link with spin glass physics, as it has been the case with constraint 
satisfaction problems\cite{zecchinaKsatScience,mezard2009information,barbier2013hard} (and actually, compressed sensing itself can also be seen as a finite temperature constraint satisfaction problem).
For instance, the compressed sensing problem can be interpreted as a densely connected (i.e. with infinite range interactions) spin glass model
of continuous spins in an external field, where the interactions enforce the state of the spins to verify the linear measurements (up to noise, interpreted as the temperature)
and the external field would be the prior in the Bayesian setting. The problem of inferring the minimum mean-squared error (MMSE) signal that generated the measurements (or ``planted'' solution in the physics language) is equivalent to sampling from the Boltzmann measure of the appropriate spin glass model, and the maximum-a-posteriori (MAP) estimate is given by the ground state of the Hamiltonian (\cite{KrzakalaPRX2012,KrzakalaMezard12} for a more detailed discussion of the links between compressed sensing and spin glass physics). Similar mappings can be established for many other computer science, inference and machine learning problems\cite{KabashimaKMSZ14}.
The typical phenomenology of spin glasses is observed in these inference problems: phase transitions and dynamical slowing down of the reconstruction algorithms near the critical ``temperature'' (the critical measurement rate in CS). Furthermore, message-passing algorithms, such as belief-propagation, can be interpreted in terms of the cavity method written for single instances, although the cavity method has been originally developed for computing thermodynamical quantities (i.e. averaged over the source of disorder) in spin glasses\cite{mezard2009information}.

The aim of this paper is to study one way to tackle very large single instances of inference problems such as compressed sensing.
To work with large signals and matrices,
however, one needs fast and memory efficient solvers. Indeed, the mere
storage of the measurement matrix in memory can be problematic as soon as the signal size $N>O(10^4)$. A classical trick (see for instance
\cite{do2008fast}) is thus to replace the random sensing matrix with a
structured one, typically random modes of a Fourier or Fourier-like
matrix. The use of the fast Fourier transform makes matrix
multiplication faster ($O(N\log N)$ instead of $O(N^2)$ operations), and
thus both speeds up the reconstruction algorithm and removes the need
to store the matrix in memory. This is also important for coding
applications where $O(N^2)$ operations can be burdensome for the
processor.

Although CS reconstruction is typically based on convex optimization
\cite{DonohoMaleki10}, we shall consider here an alternative
technique, the Approximate Message-Passing (AMP) algorithm, that
allows Bayesian reconstruction
\cite{DonohoMaleki09,DonohoMaleki10,Rangan10b,VilaSchniter11,KrzakalaPRX2012}
with much improved performances. Of special interest has been the
joint use of AMP and specially designed \cite{KrzakalaPRX2012} random
matrices based on spatial-coupling 
\cite{KudekarRichardson10,KudekarPfister10,YedlaJian12}: this has been
shown to achieve optimal information-theoretic performance in such
sparse estimation problems \cite{KrzakalaPRX2012,DonohoJavanmard11}.

Let us summarize the contributions of the present paper: while using
Fourier or Hadamard matrices has often been done with AMP (see
e.g. \cite{JavanmardMontanari12,kamilovsparse}), we provide here a
close examination of the performance of AMP with Fourier and Hadamard
operators for complex and real sparse signals respectively. As
suggested by the heuristic replica
analysis\cite{vehkapera2013analysis,wen2014analysis}, such matrices
often lead to better performances than random ones.

Secondly, inspired by the Gabor construction of
\cite{JavanmardMontanari12} ---that allowed optimal sampling of a
random signal with sparse support in frequency domain--- we
extend the construction of spatially coupled (or ``seeded''
\cite{KrzakalaPRX2012,KrzakalaMezard12}) matrices to a structured form
using fast Fourier/Hadamard operators, which allow to deal with large
signal sizes. Given the lack of theoretical guaranties, we numerically
study this strategy on synthetic problems, and compare its performance
and behavior with those obtained with random i.i.d Gaussian matrices. Our main
result is that after some randomization procedure, structured
operators appear to be nearly as efficient as random i.i.d
matrices. In fact, our empirical performances are as good as those
reported in \cite{KrzakalaPRX2012,KrzakalaMezard12} despite the
drastic improvement in computational time and memory.

Finally, to show the potential of these operators, we apply them to
sparse superposition codes, an error-correcting scheme for the
Gaussian
Channel\cite{barron2010sparse,barron2011analysis,barron2012high} for
which an AMP decoder has recently been
proposed~\cite{barbier2014replica}. We will show empirically that the
use of the spatially coupled Hadamard operator allows to get closer to
the capacity with the AMP decoder, even for moderate block-lengths.

\section{Problem setting}
\label{sec:setting}
\begin{figure}
\centering
\includegraphics[width=10cm]{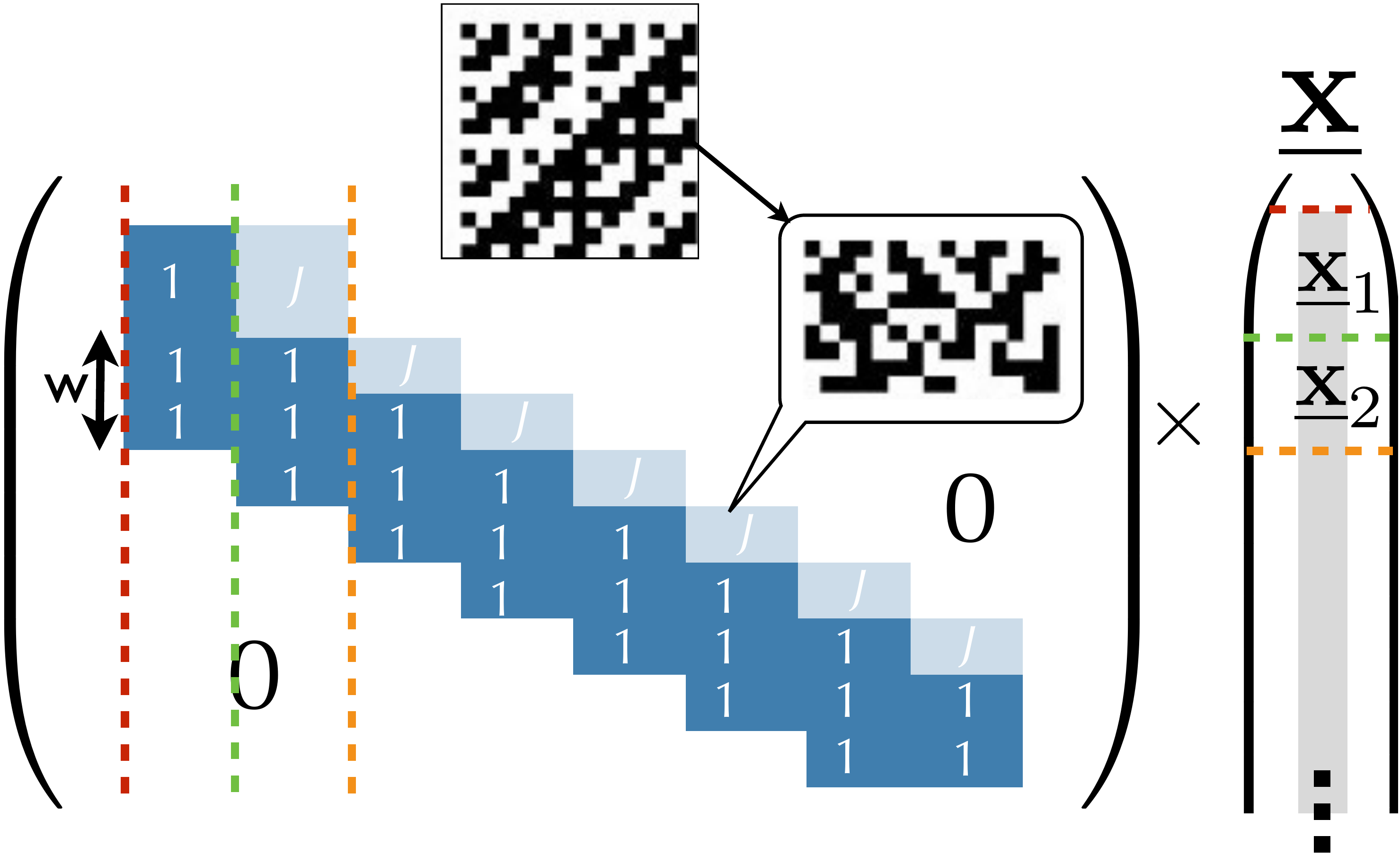}
\centering
\caption{Representation of the spatially coupled Hadamard sensing
  matrix used in our study. The operator is decomposed in $L_r\times
  L_c$ blocks, each being made of $N/L_c$ columns and $\alpha_{\rm
    seed}N/L_c$ lines for the blocks of the first block-row,
  $\alpha_{\rm rest}N/L_c$ lines for the following block-rows (these follow from the definition of $\alpha:=M/N$ combined with (\ref{eq_alpha}) ), with
  $\alpha_{\rm seed}>\alpha_{\rm rest}$. The figure shows how the
  lines of the original Hadamard matrix (of size $N/L_c\times N/L_c$)
  are randomly selected, re-ordered and sign-flipped to form a given
  block of the final operator. There is a number $w$ (the coupling
  window) of lower diagonal blocks with elements $\in \{\pm 1\}$ as
  the diagonal blocks, the upper diagonal blocks have elements $\in
  \{\pm\sqrt{J}\}$ where $\sqrt{J}$ is the coupling strength, all the
  other blocks contain only zeros. The colored dotted lines help to
  visualize the block decomposition of the signal induced by the
  operator structure: each block of the signal will be reconstructed
  at different times in the algorithm (see Fig.~\ref{fig_expVsDE} main
  figure). The procedure is exactly the same for constructing
  spatially-coupled Fourier operators, replacing the small Hadamard
  operator from which we construct the blocks by a small Fourier
  operator. The parameters that define the spatially coupled operator
  ensemble
  $(L_c,L_r,w,\sqrt{J},\alpha_{{\rm{seed}}},\alpha_{{\rm{rest}}})$
  remain the same.}
\label{fig_seededHadamard}
\end{figure}
In the following, complex variables will be underlined: $\cv{x}_j =
x_{j,1} + ix_{j,2} \in \mathbb{C}$, bold letters will be used for
vectors and capital bold letters for matrices/operators.  We will
write $x \sim \mathcal{N}_{\bar x,\sigma^2}$ if $x$ is a Gaussian
random variable with mean $\bar x$ and variance $\sigma^2$, and
$\cv{x} \sim \mathcal{C}\mathcal{N}_{\bar x,\sigma^2}$ if the real and
imaginary parts $x_1$ and $x_2$ of the random variable $\cv{x}$ are
independent and verify $x_1 \sim \mathcal{N}_{\Re{\bar x},\sigma^2}$
and $x_2 \sim \mathcal{N}_{\Im{\bar x},\sigma^2}$.  $\delta()$ is the
Dirac delta function and $\delta_{i,j}$ is the Kronecker symbol.

The generic problem we consider is as follows: An unknown signal $\cv{\bx} =
[\cv{x}_1,\cdots,\cv{x}_N]$ is passed through a linear $M \times N$
operator $\cv{\bF} = \{ \cv{F}_{\mu i}\}$, giving $\cv{\tilde \by} =
[\cv{\tilde y}_1,\cdots,\cv{\tilde y}_M]$ such that:
\begin{equation}
\label{eq_tildeY}
\cv{\tilde y}_{\mu} = \sum_{i=1}^N \cv{F}_{\mu i} \cv{x}_{i}.
\end{equation}
In the most general setting, a noisy version $\cv{\by} =
[\cv{y}_1,\cdots,\cv{y}_M ]$ of $\cv{\tilde \by}$ is then measured:
\begin{equation}
\label{eq_Y}
\cv{\by} = \cv{\tilde \by} + \cv{\bxii}, \quad \cv{\bxii}=[\cv{\xi}_1,\cdots,\cv{\xi}_M],
\end{equation}  
with $\cv{\xi}_{\mu} \sim \mathcal{C}\mathcal{N}_{0,\Delta}\ \forall \
\mu \in \{1,\cdots,M\}$.  The goal is to recover the original vector
from the noisy measurements.

We shall consider two particular settings of this problem in the present
paper. The first one is noiseless compressed sensing with real and complex
variables. We will use the following Gauss-Bernoulli distribution to
generate $\rho$-sparse complex random vectors:
\begin{equation}
\label{Px}
P(\cv{\bx}) =\prod_{j=1}^N \left[(1 - \rho) \delta(\cv{x}_j) + \rho \ \mathcal{C}\mathcal{N}_{\bar x,\sigma^2}(\cv{x}_{j})\right].
\end{equation}
Here we shall assume that the correct values for $\rho$, $\bar{x}$, $\sigma^2$ and $\Delta$,
as well as the empirical signal distribution (\ref{Px}) are known. As
shown empirically in
\cite{VilaSchniter11,KrzakalaPRX2012,KrzakalaMezard12} these
parameters can be learned efficiently with an expectation
maximization procedure if unknown. A remarkable result is that if
$\cv{\bx}$ is sparse, i.e. has a fraction $(1-\rho)$ of strictly zero
components, then it is possible to recover $\cv{\bx}$ from the
knowledge of $\cv{\by}$ and $\cv{\bF}$ even if the system is
underdetermined, i.e. the measurement rate $\alpha := M/N <1$. An
information-theoretical lower bound for the measurement rate is
$\alpha_{\rm min}(\rho) = \rho$ : no algorithm can perform a
reconstruction below this bound. CS algorithms do not usually reach
this bound, but are characterized by a phase transition that depends
on the sparsity and the noise levels $\alpha_{\rm
  PT}(\rho,\Delta)>\rho$, above (below) which the algorithm
succeeds (fails) with high probability (i.e. that tends to one as the
signal size $N\to\infty$).

In the last part of the paper, we will consider sparse superposition
codes. In that case, we will denote the length of vector $\bx$ as
$L$ and each of its components will be a $B$-dimensional real
variable. The prior, for this problem, will enforce the condition that
each of the $B$-dimensional variables is pointing towards one of the
summits of the $B$-dimensional hypercube, i.e. only one component among the $B$ is $1$, the others are $0$. Such codes are, in a proper
limit, capacity achieving \cite{barron2010sparse}. However, in the
standard version of the problem, AMP decoding is limited to a region
far from capacity~\cite{barbier2014replica}. We will come back to this
problem in section \ref{Sec:super} and will first concentrate on noiseless
compressed sensing.

\subsection{Spatially coupled measurement matrices}
AMP is a CS solver that can perform
Bayes-optimal inference, and the position of its phase transition
$\alpha_{\rm AMP}(\rho)$ can be determined exactly in the $N \to
\infty$ limit by State Evolution (SE) analysis
\cite{BayatiMontanari10}.  Though it was initially designed for i.i.d
random matrices, it was recently empirically shown, through the use
of the replica method \cite{KrzakalaMezard12}, and then, also, theoretically
proven \cite{DonohoJavanmard11} that the use of 
spatially coupled matrices allows to lower the phase transition from
$\alpha_{\rm AMP}(\rho)>\rho$ to $\alpha_{\min}(\rho)=\rho$.

Fig.\ref{fig_seededHadamard} shows such a spatially coupled matrix:
It has a block structure with $L_r\times L_c$ blocks, each block containing either
only zeros or a different random selection of modes of a Fourier or Hadamard
operator. Each of these blocks is constructed from the same original
operator of size $N/L_c\times N/L_c$ and the differences from one block to another arise from
the selected modes, their permutation and signs that are randomly changed.
One of the blocks is called the seed (usually the block on the upper left corner as shown on Fig.~\ref{fig_seededHadamard}) and its ratio of number of lines over number of columns $\alpha_{\rm seed}$ 
has to be larger than $\alpha_{\rm AMP}(\rho)$ for the spatial coupling to work.
All the other blocks can have their
$\alpha_{\rm{rest}}$ asymptotically as low as $\rho$.  The first
block of the signal is then easily reconstructed due to its high
$\alpha_{\rm{seed}}$, and the solution spreads through the
entire signal thanks to the coupling introduced by the non-diagonal
blocks. This ``nucleation'' effect (that has the same phenomenology than the surfusion of super-cooled water for example) is discussed in full details in
\cite{KudekarRichardson10, KrzakalaMezard12}.

The link between the overall measurement rate $\alpha$, that of the seed $\alpha_{{\rm{seed}}}$ and that of the bulk $\alpha_{{\rm{rest}}}$ is given by:
\begin{equation}
\alpha_{{\rm{rest}}}= \frac{\alpha L_c -\alpha_{{\rm{seed}}}}{L_r - 1} = \alpha\frac{L_c - \beta_{{\rm{seed}}}}{L_r - 1}.
\label{eq_alpha}
\end{equation}
In practice, $\alpha$ is fixed and $\alpha_{{\rm{seed}}}:=\alpha\beta_{{\rm{seed}}}$ as well by fixing $\beta_{{\rm{seed}}}$. $\alpha_{{\rm{rest}}}$ is then deduced from (\ref{eq_alpha}). 
In the rest of the paper, we will define the spatially coupled ensemble by $(L_c,L_r,w,\sqrt{J},\alpha, \beta_{{\rm{seed}}})$ instead of $(L_c,L_r,w,\sqrt{J},\alpha_{{\rm{seed}}},\alpha_{{\rm{rest}}})$.
\subsection{The AMP algorithm}
We now describe the AMP algorithm for complex signals (cancelling all the imaginary parts, AMP for real signals is recovered). 
AMP is an iterative algorithm that calculates successive estimates of $\cv{\bx}$ and $\cv{\tilde \by}$, 
as well as uncertainties on those estimates.
It uses both linear steps, involving matrix multiplications, and non-linear steps, involving thresholding functions 
$\cv{f}_{a_i}$ and ${f}_{c_i}$ (the underline means that $\cv{f}_{a_i}$ outputs a complex number). These are either calculated from the actual signal distribution $P(\cv{\bx})$, in which case the algorithm is Bayes-optimal, or from a sparsity-inducing Laplace prior, in which case AMP 
solves an $L_1$ minimization problem. 

In order to avoid confusion with the literature where variations of
AMP are already presented, we will refer to the Bayes-optimal AMP by
``BP'' and ``c-BP'' for the real and the complex case respectively,
and to the $L_1$-minimizing version by ``LASSO'' and ``c-LASSO''
respectively (where BP stands for Belief Propagation, of which AMP is an adaptation). As the thresholding functions are applied
componentwise, the time-consuming part of the algorithm is the matrix
multiplications in the linear step. Here, we use Fourier and Hadamard
operators in order to reduce the complexity of the matrix
multiplications from $O(N^2)$ to $O(N \log N)$.  The authors of
\cite{JavanmardMontanari12} have used a related, yet different way to
create spatially coupled matrices using a set of Gabor transforms.

We
define $\textbf{e}_c$, with $c\in\{1,\cdots,L_c\}$, a vector of size $N_c=N/L_c$, as the $c^{th}$ block
of $\textbf{e}$ (of size $N$) and $\textbf{f}_r$, with $r\in\{1,\cdots,L_r\}$, a vector of size
$N_r=\alpha_rN/L_c$ as the $r^{th}$ block of $\textbf{f}$ (of size
$M$). For example, in Fig.~\ref{fig_seededHadamard}, the signal
$\cv{\bx}$ is decomposed as $[\cv{\bx}_1,\cdots,\cv{\bx}_{L_c}]$. The notation $\{i\in c\}$ (resp. $\{\mu\in r\}$) means all the components of $\textbf{e}$ that are in the $c^{th}$ block of $\textbf{e}$ (resp. all the components of $\textbf{f}$ that are in the $r^{th}$ block of $\textbf{f}$). The
algorithm requires four different operators performing the following
operations:
\begin{eqnarray}
&\tilde O_\mu(\textbf{e}_c) := \sum_{\{i\in c\}}^{N/L_c} |\cv{F}_{\mu i}|^{2} e_i, \ \ \ \  O_\mu(\textbf{e}_c) := \sum_{\{i\in c\}}^{N/L_c} \cv{F}_{\mu i} e_i ,\nonumber\\
&\tilde O_i(\textbf{f}_r) := \sum_{\{\mu\in r\}}^{\alpha_rN/L_c} |\cv{F}_{\mu i}|^2 f_\mu, \ O_i^*(\textbf{f}_r) := \sum_{\{\mu\in r\}}^{\alpha_rN/L_c}\cv{F}_{\mu i}^* f_\mu .\nonumber
\end{eqnarray}
$\alpha_r$ is
the measurement rate of all the blocks at the $r^{th}$ block-row, for example in Fig.~\ref{fig_seededHadamard}, $\alpha_1 = \alpha_{{\rm{seed}}}$ and $\alpha_{j} = \alpha_{{\rm{rest}}} \ \forall \ j>1$ and ${F}_{\mu i}^*$ is the complex conjugate of ${F}_{\mu i}$. 
Because the value of $|\cv{F}_{\mu i}|^2$ is either $0$, $1$ or $J$ $\forall \ (\mu,i)$ as we use Hadamard or Fourier operators (it can be read on Fig.~\ref{fig_seededHadamard}), all these operators 
do not require matrix multiplications as they are implemented as fast transforms ($O_\mu()$ and $O_i^*()$)  or simple sums ($\tilde O_\mu()$ and $\tilde O_i()$). It results in the updates for AMP \cite{Schniter2012compressive}, with a generic operator, see Fig.~\ref{algo_AMP}.
\begin{figure}[!t] 
\centering
\begin{algorithmic}[1]
\State $t\gets 0$
\State $\delta_{\rm max} \gets \epsilon + 1$
\While{$t<t_{\rm max} \ \textbf
{and } \delta_{\rm max}>\epsilon$} 
\State $\Theta^{t+1}_\mu \gets \sum_{c=1}^{L_c}\tilde O_\mu(\textbf{v}_c^t)$
\State $\underline{{\rm w}}^{t+1}_\mu \gets \sum_{c=1}^{L_c}O_\mu(\underline{\textbf{a}}_c^t) - \Theta^{t+1}_\mu\frac{\underline{y}_\mu-\underline{{\rm w}}^t_\mu}{\Delta + \Theta^t_\mu}$
\State $\Sigma^{t+1}_i \gets \left[\sum_{r=1}^{L_r}\tilde O_i\left([\Delta + \boldsymbol{\Theta}_r^{t+1}]^{-1}\right)\right]^{-1/2}$
\State $\underline{R}^{t+1}_i \gets \underline{a}^t_i + (\Sigma^{t+1}_i)^2 \sum_{r=1}^{L_r} O_i^*\left(\frac{\underline{\textbf{y}}_r - \underline{\textbf{w}}^{t+1}_r}{\Delta + \boldsymbol{\Theta}^{t+1}_r}\right)$
\State $v^{t+1}_i \gets f_{c_i}\left((\Sigma^{t+1}_i)^2,\underline{R}_i^{t+1}\right)$
\State $\underline{a}^{t+1}_i \gets \cv{f}_{a_i}\left((\Sigma^{t+1}_i)^2,\underline{R}_i^{t+1}\right)$
\State $t \gets t+1$
\State $\delta_{\rm max} \gets \textrm{max}_i\left(|\underline{a}_i^t - \underline{a}_i^{t-1}|^2\right)$
\EndWhile
\State \textbf{return} $\{a_i\}$
\end{algorithmic}
\centering
\caption{The AMP algorithm written with operators. Depending on whether it is used on a real or complex signal, with 
Bayes-optimal or sparsity-inducing thresholding functions $\cv{f}_{a_i}$ and $f_{c_i}$, we call it BP, c-BP, LASSO or c-LASSO.
$\epsilon$ is the accuracy for convergence and $t_{\rm max}$ the maximum number of iterations.
A suitable initialization for the quantities is ($\cv{a}_i^{t=0}=0$, $v_i^{t=0}=\rho \sigma^2 $, $\cv{{\rm w}}_\mu^{t=0}=\cv{y}_\mu$). 
Once the algorithm has converged, i.e.  the quantities do not change anymore from iteration to iteration, the estimate of the $i^{th}$ signal component is $\cv{a}^{t}_i$.
The variable $\cv{\textbf{w}}$ is an estimate of $\cv{\tilde \by}$, while $\cv{\bf R}$ and $\cv{\bf a}$ are estimates of $\cv{\bx}$; $\boldsymbol{\Theta}$,
$\bxigma$ and $\bv$ are their respective uncertainties.
The nonlinear thresholding functions $\cv{f}_{a_i}$ and $f_{c_i}$ take into account the prior distribution $P(\cv{\bx})$.
 In the case of compressed sensing, applying $\cv{f}_{a_i}$ to a $\underline{R}_i^{t+1}$ close to zero will give a result even closer to zero,
  while bigger inputs will be left nearly unchanged, thus favoring sparse solutions.}  
\label{algo_AMP}
\end{figure}
%
%To avoid confusion with the litterature where variations of AMP are
%already presented, we will refer to this algorithm as c-BP; and to its
%real-only counterpart as BP (where BP stands for Belief Propagation, which AMP is just an adaptation of). 

Here, we give the functions
$\cv{f}_{a_i}$ and $f_{c_i}$ that are calculated from
$P(\cv{\bx})$ and are thus Bayes-optimal, which is not the case for
LASSO and c-LASSO \cite{maleki2012asymptotic}.
For BP, they are
given in \cite{KrzakalaMezard12,BarbierKrzakalaAllerton2012}. 
For c-BP, the signal is complex and drawn from the distribution~(\ref{Px}), and 
the thresholding functions (which give posterior scalarwise estimates of the mean and variance) are given by:
\begin{eqnarray}
\cv{f}_{a_i}(\Sigma^2,\cv{R}) &= g\rho\chi^2\cv{M}/Z ,\nonumber\\
f_{c_i}(\Sigma^2,\cv{R}) &= \left(g\rho\chi^2 \(|\cv{M}|^2 + 2 \chi^2\)/Z - |\cv{f}_a(\Sigma^2,\cv{R})|^2\right)/2 ,\nonumber
\end{eqnarray}
\noindent
with the following definitions:
\begin{eqnarray}
&\cv{M} := (\sigma^2 \cv{R}+\Sigma^2 \bar x)/(\Sigma^2+\sigma^2), \ \chi^2 := \Sigma^2\sigma^2/(\Sigma^2 + \sigma^2), \nonumber \\
&g := e^{-\frac{1}{2}\left(\frac{|\bar x|^2}{\sigma^2} + \frac{|\cv{R}|^2}{\Sigma^2} - \frac{|\cv{M}|^2}{\chi^2}\right)},\ Z := \sigma^2(1-\rho)e^{-\frac{|\cv{R}|^2}{2\Sigma^2}} + \rho\chi^2 g\nonumber,
\end{eqnarray}
where $\cv{R}$ and $\Sigma$ are defined in algorithm~\ref{algo_AMP}.
These functions are not identical to the ones for the real case since in the
prior distribution (\ref{Px}), the real and imaginary parts of the
signal are jointly sparse (i.e. have same support but independent values), which can be a good assumption, for
instance in MRI.  As in c-LASSO~\cite{maleki2012asymptotic}, this allows to lower the phase transition
compared to when the real and imaginary part of the signal are
assumed to be independent.

The implementation requires caution: the necessary ``structure killing'' randomization 
is obtained by applying a permutation of lines after the use of the fast operator.
For each block $(r,c)$, we choose a random subset of modes $\Omega^{r,c} = \{ \Omega^{r,c}_1 ,\cdots, \Omega^{r,c}_{N_r} \} \subset \{ 1,\cdots, N_c \}$.
The definition of 
$O_\mu(\textbf{e}_c)$ using a standard fast transform ${\rm FT}$ will be:
\begin{equation}
 O_\mu(\textbf{e}_c) :=   {\rm FT}(\textbf{e}_c) \mid_{\Omega^{r_{\mu},c}_{\mu - \mu_{r_{\mu}} + 1}}
\end{equation}
where $r_{\mu}$ is the index of the block row that includes $\mu$, $\mu_{r_{\mu}}$ is the number of the first line of the block row $r_{\mu}$ and $\lambda \mid_{\mu}$ is the $\mu^{\rm th}$ component of $\lambda$.
 For $O_i^*(\textbf{f}_r)$ instead,
\begin{equation}
 O_i^*(\textbf{f}_r):={\rm FT}^{-1}( \tilde{\textbf{f}}_r ) \mid_{i - i_{c_i}+1}
\end{equation}
where $c_{i}$ is the index of the block column that includes $i$, $i_{c_{i}}$ is the number of the first column of the block column $c_{i}$,
${\rm FT}^{-1}$ is the standard fast inverse operator of ${\rm FT}$ and $\tilde{\textbf{f}}_r$ is defined in the following way:
\begin{equation}
  \forall \gamma \in \{ 1,\cdots,N_r \}, \quad \tilde{\textbf{f}}_r \mid_{\Omega_{\gamma}^{r,c}} = \textbf{f}_r \mid_{\gamma} \quad {\rm and} \quad  \forall i \notin \Omega^{r,c}, \, \tilde{\textbf{f}}_r \mid_i = 0. 
\end{equation}
The mean squared error (MSE) achieved by the algorithm is: 
\begin{equation}
E^t:=||\cv{\boldsymbol{a}}^t - \cv{\bx}||_2^2=1/N\sum_{i=1}^N |\cv{a}_i^t - \cv{x}_i|^2,
\label{eq_MSE}
\end{equation}
and measures how well the signal is reconstructed.
\begin{figure}[!t]
\centering
\includegraphics[width=0.6\textwidth]{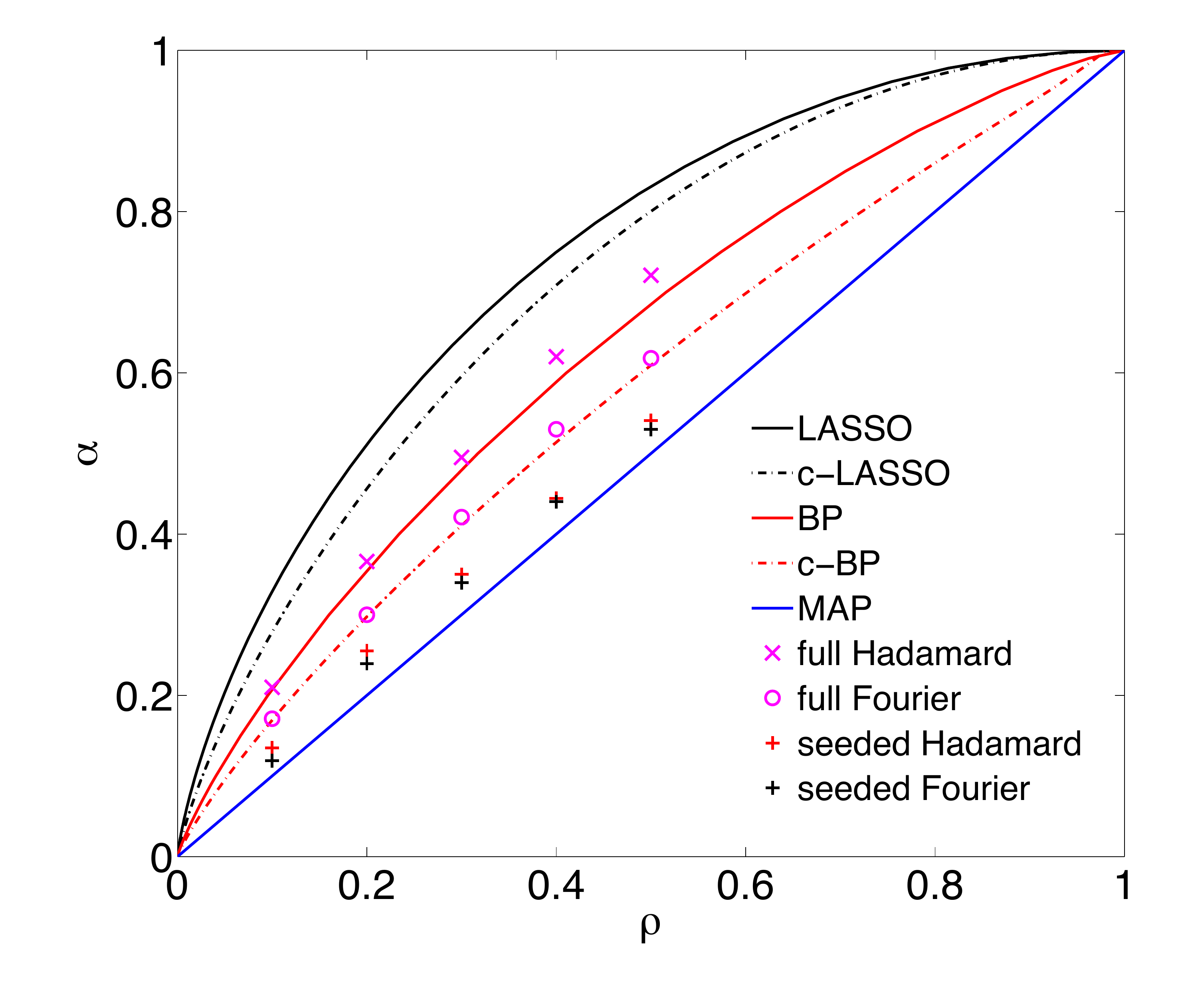}
\caption{Phase diagram on the $\alpha=M/N$ (measurement rate) vs $\rho$
(sparsity rate) plane in the noiseless case $\Delta=0$. Lines are phase transitions predicted by the
state evolution technique for i.i.d random Gaussian matrices, while markers are
points from experiments using structured operators with empirically optimized parameters. Good sets of parameters usually lie in the following sets: $(L_c\in\{8,16,32,64\}, L_r = L_c + \{1,2\}, w\in\{2,\cdots,5\}, \sqrt{J}\in[0.2,0.7],\beta_{{\rm{seed}}}\in[1.2,2])$. With larger signals, higher values of $L_c$ are better.
Just as c-LASSO
allows to improve the usual LASSO phase transition when the complex
signal is sampled according to (\ref{Px}) (thanks to the joint
sparsity of the real and imaginary parts), c-BP improves the usual
BP transition.  The line $\alpha=\rho$ is both the maximum-a-posteriori (MAP) threshold for
CS and the (asymptotic) phase transition with spatially coupled
matrices. Pink experimental points correspond to perfectly reconstructed instances
using non spatially coupled Hadamard and Fourier operators (on the BP and c-BP
phase transition respectively), the black and red points to spatially coupled ones (close to the MAP
threshold). Properly randomized structured operators appear to have
similar performances as random measurement matrices.}

\label{fig_DE}
\end{figure}
\section{Results for compressed sensing}
\label{sec:results}
\begin{figure}
\centering
\includegraphics[width=0.8\textwidth]{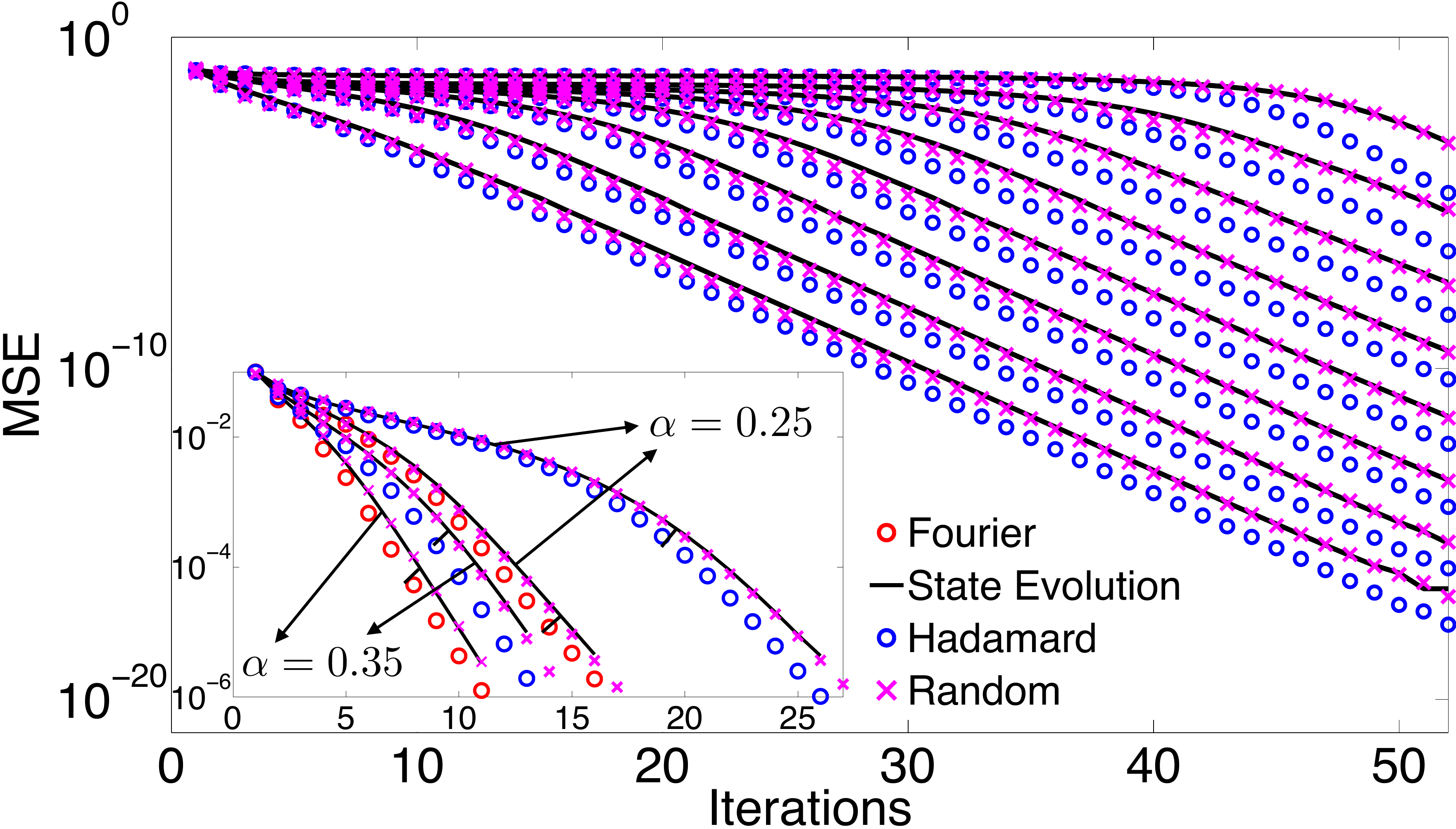}
\caption{Comparison of the mean squarred-error predictions between the SE (black lines) and the actual
behavior of the algorithm for spatially coupled matrices (main figure) and standard
ones (inset) both with structured operators (circles) and random i.i.d
Gaussian matrices (crosses) in a noiseless conpressed sensing problem. In both plots, the signal size is $N= 2^{14}$
  with the random i.i.d Gaussian matrices, and $N=  2^{20}$ %\approx10^{6}$ 
  with the operators and are generated
  with ($\rho= 0.1$, $\bar x= 0$, $\sigma^2=
  1$, $\Delta=0$). While experiments made with random i.i.d
  matrices fit very well the SE predictions, those with
  the structured operators are not described well by the SE, although
  final performances are comparable. \textbf{Main:} For an Hadamard spatially coupled matrix as in
  Fig.~\ref{fig_seededHadamard} with ($L_c=  8$, $L_r = L_c+2$, 
  $w = 1$, $\sqrt{J} = 0.1$, $\alpha = 0.22$, $\beta_{{\rm{seed}}} = 1.36$). Each curve corresponds
  to the MSE tracked in a different block of the real signal \bx \
  (see Fig.~\ref{fig_seededHadamard}). \textbf{Inset:} Reconstructions made
  with standard non spatially coupled (or full) matrices at $\alpha = 0.35$ and $\alpha = 0.25$. The reconstruction
  with the Fourier operator of a complex signal (instead of real with Hadamard) is faster thanks to the joint
  sparsity assumption of (\ref{Px}). The arrows identify
  the groups of curves corresponding to same measurement rate $\alpha$. Both in the Fourier and
  Hadamard cases, we observe that convergence is slightly faster than
  predicted by the SE analysis.}
  \label{fig_expVsDE}
  
  \end{figure}
When the sensing matrix is i.i.d random, or spatially coupled with i.i.d random blocks, the evolution of $E^t$
in AMP can  be predicted in the large signal limit on
a rigorous basis called state evolution \cite{BayatiMontanari10,DonohoJavanmard11,bayati2012universality}.
For BP with real Gauss-Bernoulli signals, this analysis  can be found in
\cite{KrzakalaMezard12}. For c-BP with i.i.d Gaussian matrices, the derivation goes
very much along the same lines and we shall report the results
briefly. The evolution of $E^t$ is given by the following equation:
\begin{equation}
E^{t+1} = \int \mathcal{D}\underline z \left[(1-\rho) f_{c_i} \left((\Sigma^t)^2, \cv{R}_1^t(\underline z)\right)+\rho f_{c_i} \left((\Sigma^t)^2, \cv{R}_2^t(\underline z) \right)\right], \nonumber
\label{complexDE}
\end{equation}
where:
\begin{eqnarray}
&\underline z := z_1 + i z_2, \ (\Sigma^t)^2:=(\Delta+E^{t})/\alpha ,\nonumber\\
&\cv{R}_u^t(\underline z):=\underline z\sqrt{\sigma^2\delta_{u,2} + (\Sigma^t)^2}\nonumber,\ \mathcal{D}\underline z := dz_1dz_2 \frac{e^{-\frac{1}{2} (z_1^2 + z_2^2)}}{2 \pi}.
\end{eqnarray}
Note that this SE equation is the same as given in \cite{maleki2012asymptotic},
despite slightly different update rules in the algorithm. 

For c-BP with spatially coupled matrices with i.i.d Gaussian blocks, the expression involves the MSE in each block $p\in\{1,\cdots,L_c\}$ (see main
figure of Fig.~\ref{fig_expVsDE}), and becomes \cite{KrzakalaMezard12}:
\begin{eqnarray}
\label{densEvo}
E_p^{t+1} &=&\int \mathcal{D}\underline z \left[(1-\rho)f_c\((\Sigma^t_p)^2, \cv{R}_{p,1}^t(\underline z)\) +\rho f_c\((\Sigma^t_p)^2, \cv{R}_{p,2}^t(\underline z)\)\right] , \nonumber
\end{eqnarray}
where:
\begin{eqnarray}
&(\Sigma^t_p)^2=\left[n_p \sum_{q=1}^{L_r} \frac{ \alpha_{q} J_{qp}}{\Delta + \sum_{r=1}^{L_c} n_{r} J_{qr}E_r^{t}}\right]^{-1}, \nonumber\\ 
%&\cv{R}_{p,u}^t(\underline z)=\underline z \left((\Sigma^t_p)^{-2} + \delta_{u,2} \right)^{-\frac{1}{2}} \nonumber ,
&\cv{R}_{p,u}^t(\underline z)=\underline z \sqrt{\sigma^2 \delta_{u,2} + (\Sigma^t_p)^{2}} \nonumber ,
\end{eqnarray}
where $n_{i} := N_{i} / N =1/L_c$ for matrices with equally wide blocks (i.e. $N_{i}=N/L_c\ \forall \ i \in \{1,\dots,L_c\}$) as in Fig.~\ref{fig_seededHadamard}, $\alpha_{k}=\alpha_{\rm rest} + (\alpha_{\rm seed} - \alpha_{\rm rest})\delta_{k,1}$ and $J_{qp}$ is the variance of the elements belonging to the block at the $q^{th}$ block row and $p^{th}$ block column (1 or $J$ in Fig.~\ref{fig_seededHadamard}).
\begin{figure}[!t]
\centering
\includegraphics[width=0.8\textwidth]{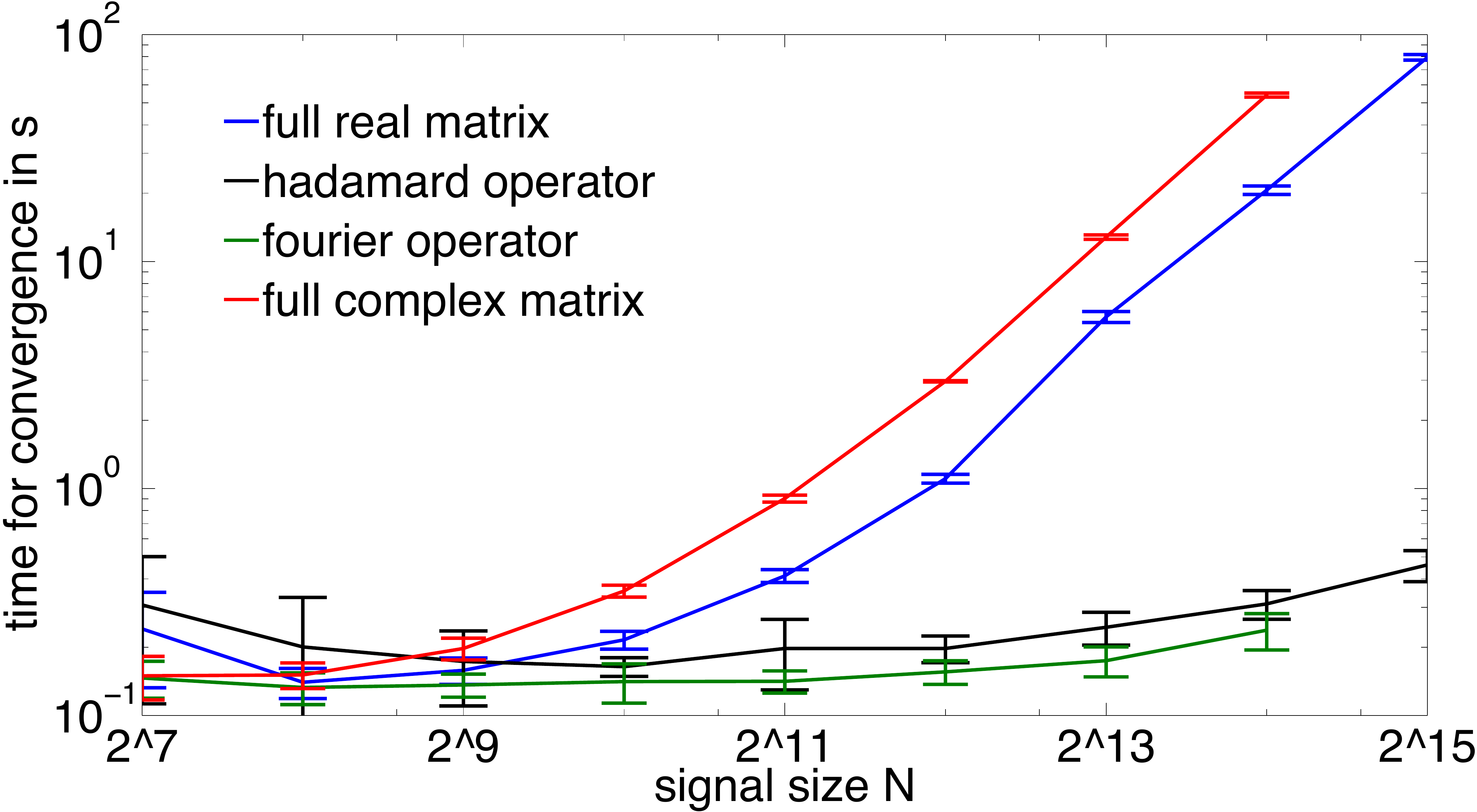}
\caption{Time required for convergence (i.e. MSE $< 10^{-6}$) of the AMP algorithm in seconds as a function of the signal size, in the non spatially coupled case for a typical compressed sensing problem.
The signal has distribution given by (\ref{Px}) and is real (complex) for the reconstruction with real (complex) matrices. 
The plot compares the speed of AMP with matrices (blue and red lines) to those of AMP using the structured operators (black and green lines). 
The points have been averaged over 10 random instances and the error bars represent the standard deviation with respect to these. 
The simulations have been performed on a personal laptop. As the signal size increases, the advantage of using operators becomes obvious.}
\label{fig_timeComp}

\end{figure}

We now move to our main point. In the case of AMP with structured (Fourier or Hadamard) operators instead of i.i.d matrices,
the SE analysis cannot be made. Hence we experimentally compare the performances between
AMP with structured operators and i.i.d matrices. The comparison is shown in
Fig.~\ref{fig_expVsDE}, which opposes theoretical results from SE and experimental results obtained by
running AMP on finite-sized data. On Fig.~\ref{fig_DE}, we
show the phase transition lines obtained by SE analysis in the $(\alpha,\rho)$
plane, and we added markers showing the position of instances actually recovered by the algorithm with spatially coupled structured operators in the noiseless case $\Delta=0$.

\subsection{Full operators}

Let us first concentrate on AMP with non spatially coupled (or full) structured operators. The
first observation is that the SE {\it does not} correctly describe the
evolution of the MSE for AMP with full structured operators (inset Fig.~\ref{fig_expVsDE}).
It is perhaps not surprising, given that AMP has
been derived for i.i.d matrices. The difference is
small, but clear: $E^t$ decreases faster with structured operators than with i.i.d matrices.
 However, despite this slight difference in the dynamical behavior of the algorithm,
the phase transitions
and the final MSE performances for both approaches appear to be
extremely close. As seen in Fig.~\ref{fig_DE}, for small
$\rho$, we cannot
distinguish the actual phase transition with structured operators from the one predicted by SE. 
Thus, the SE analysis is still a good tool to predict the performances 
of AMP with structured operators.

\subsection{Spatially coupled operators}
For spatially coupled operators, the conclusions are similar (main plot on
Fig.~\ref{fig_expVsDE}). Again, $E_p^t$ (in each of the blocks of the signal induced by the spatially coupled structure of the measurement matrix) decreases faster with structured operators than with i.i.d matrices. 
But our empirical results are
consistent (see Fig.~\ref{fig_DE}) with the hypothesis that the
proposed scheme, using spatially coupled Fourier/Hadamard operators, achieves correct
reconstruction as soon as $\alpha > \rho$ when $N$ is large. Indeed, we observe that the
gap to the MAP threshold $\alpha_{\rm min}=\rho$ decreases as the signal size increases upon
optimization of the spatially coupled operator structure. The results in
Fig.~\ref{fig_DE} and Fig.~\ref{fig_expVsDE} are obtained with spatially coupled
matrices of the ensemble: $(L_c=8, L_r = L_c + 1, w \in\{1,2\}, \sqrt{J} \in \[0.2, 0.5\], \beta_{\rm seed} = [1.2,1.6])$. While these parameters do not
quite saturate the bound $\alpha=\rho$ (which is only possible for $L_c
\to \infty$
\cite{KudekarRichardson10,KrzakalaPRX2012,DonohoJavanmard11}), they
do achieve near optimal performances.
This, as well as the substantial cut in running time (Fig.~\ref{fig_timeComp}) 
with respect to AMP with i.i.d matrices and the possibility to work with very large systems without saturating the memory, strongly supports the advantages of the proposed implementation of AMP.

% \subsection{Influence of noise}
% The results in Fig.~\ref{fig_expVsDE} are presented for noiseless data. 
% In the presence of a noise with variance $\Delta$, compressed sensing with full operators 
% is still well described by the density evolution equation~(\ref{complexDE}), that takes into account the noise. 
% The evolution of the MSE is qualitatively the same as without noise, with the difference that the MSE does not decrease 
% below a level that is of the order of $\Delta$, thus reaching a plateau~\cite{barbier2012compressed}.
% In contrast, in the case of spatially coupled operators, experimental results agree less well with the density evolution: 
% AMP with spatial coupling does not seem to be very robust to noise, leading to high final MSEs even for low noise levels.
% % 
% % \begin{figure}
% %  \centering
% % % \includegraphics[width=0.8\textwidth]{}
% %  \label{Achievable MSE for different noise levels.}
% % \end{figure}

\section{Application to superposition codes}
\label{Sec:super}
\begin{figure}
\centering
\begin{tabular}{cc}
  \includegraphics[width=8cm]{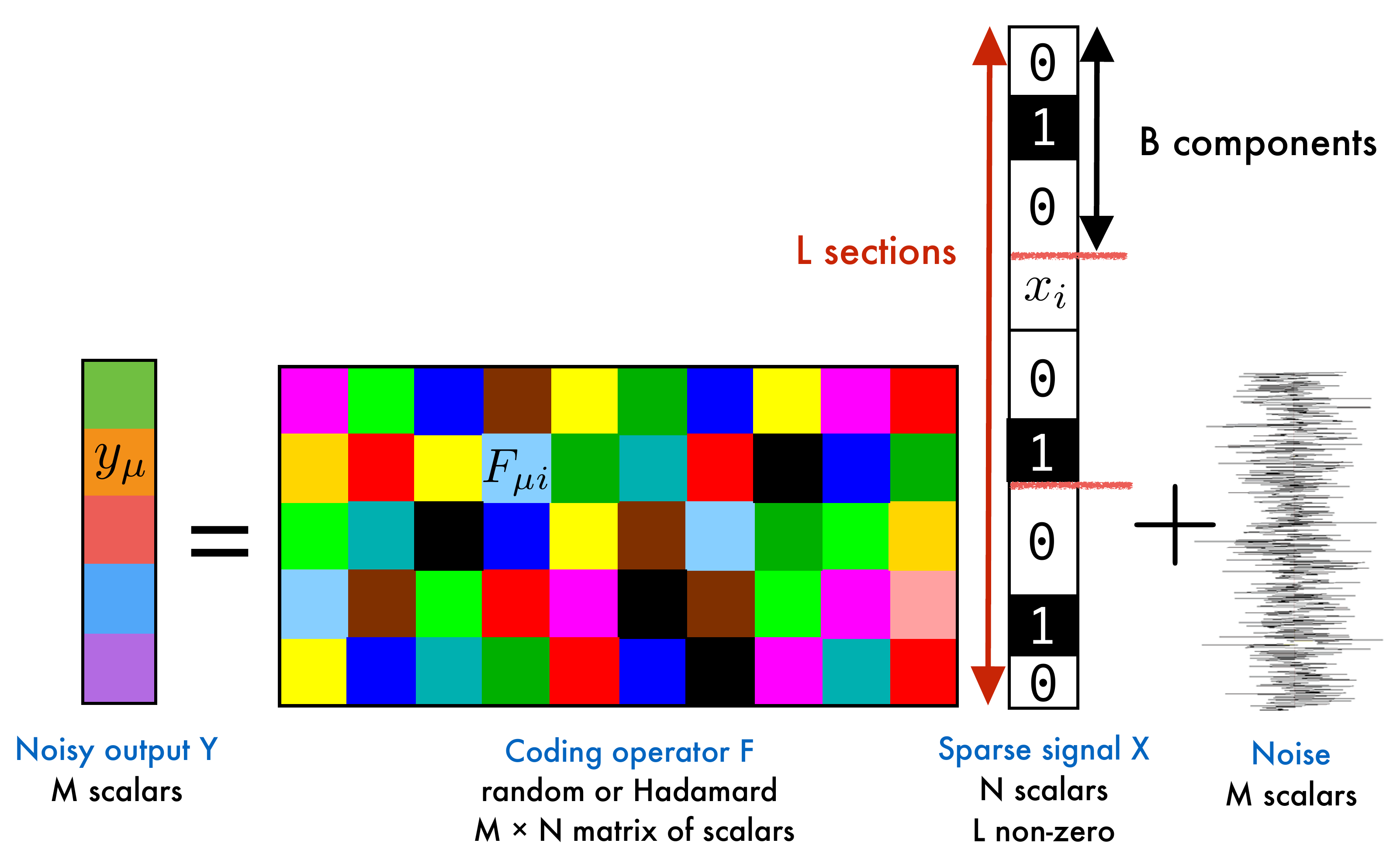}
  \includegraphics[width=8cm]{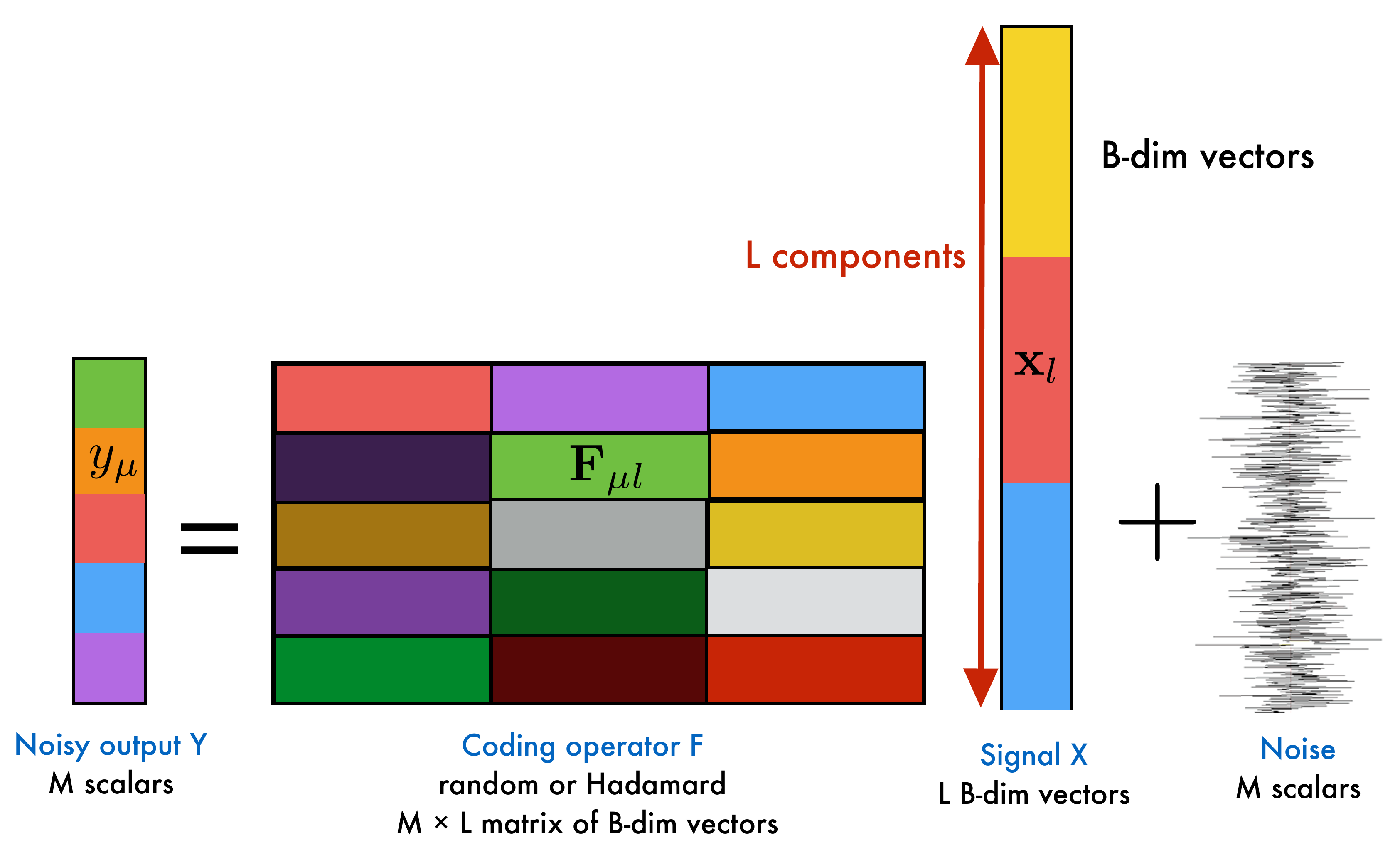}
\end{tabular}
\caption{$\underline{\textbf{Left}:}$ Representation of the ``compressed sensing like'' problem associated to the decoding of the sparse signal over the AWGN channel, in terms of the usual $1$-dimensional variables. In this setting, the variables in the same section $\{x_i : i \in l\}$ are strongly correlated due to the hard constraint that only one of them can be non-zero. The matrix elements are scalars as the signal components.
$\underline{\textbf{Right}:}$ Re-interpretation of the decoding in terms of $B$-dimensional variables. In this version, the matrix elements are grouped to form $B$-dimensional vectors that are applied (using the usual scalar product for vectors) on the associated $B$-dimensional vectors representing the new components of the signal. The advantage of this setting is that now, the signal elements are uncorrelated, which is useful in the Bayesian setting to write a factorized prior distribution.}
\label{fig_1dOp}

\end{figure}
We shall now show how the present construction can be applied to a
practical problem of coding theory: the sparse superposition codes
introduced in
\cite{barron2010sparse,barron2011analysis,barron2012high} for error
correction over the Gaussian channel.  The AMP decoder for this problem
has been studied in \cite{barbier2014replica}.  While these codes are
capacity acheving \cite{barron2011analysis}, the AMP decoder is not
\cite{barbier2014replica} (at least without a proper power
allocation). Our aim here is to show numerically that the
spatially coupled Hadamard operator is appropriate to transmit close
to the capacity, much in the same way spatial-coupling has been
applied to Low Density Parity Check codes (LDPC)
\cite{kudekar2011threshold}.

\subsection{Superposition codes}
\begin{figure}
\centering
\includegraphics[width=12cm]{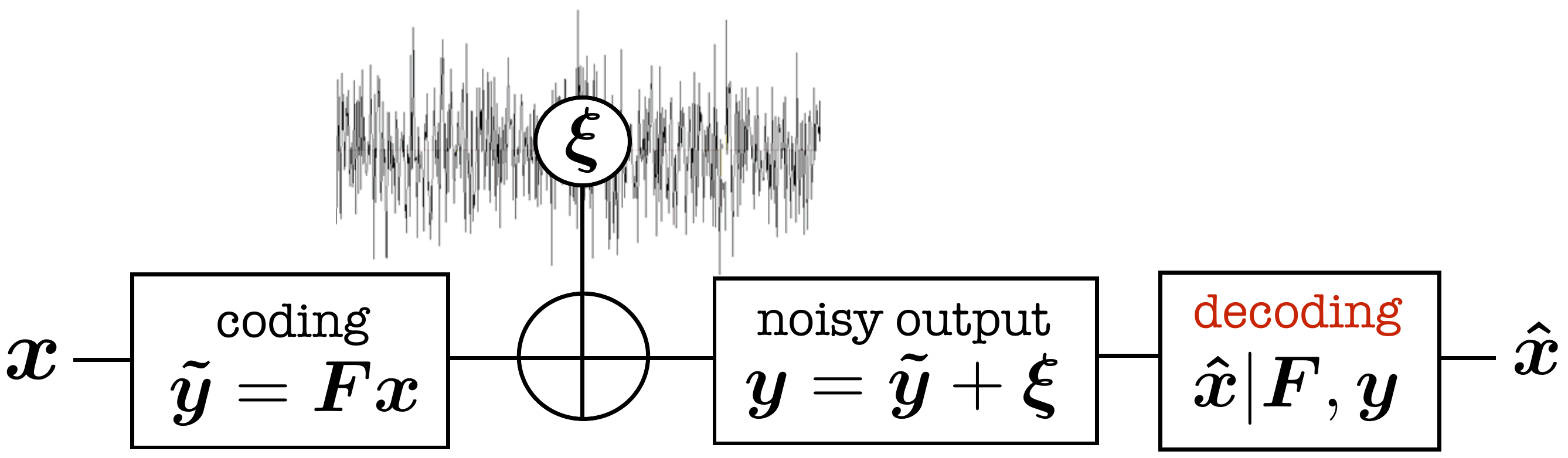}
\caption{AWGN channel model: the message $\bx$ is coded by a linear transform, $\tilde \by = \bF\bx$ and then the real encoded message $\tilde \by$ is sent through the AWGN channel that adds an i.i.d Gaussian noise $\bxii$ with zero mean and a given variance $\Delta$. The receptor gets the corrupted message $\by$ and outputs an estimate $\hat \bx$ of $\bx$, knowing $\bF$ and $\by$. Perfect decoding means that $\hat \bx = \bx$.}
\label{fig_AWGN}

\end{figure}

Let us first give the basics of sparse superposition codes. Our goal is to
transmit a real message ${\tilde \bx} = \{\tilde x_i : i\in
\{1,...,L\}, \tilde x_i \in \{1,...,B\} \ \forall \ i\}$ of size $L$
through an Additive White Gaussian Noise (AWGN) channel, where a
Gaussian noise with zero mean and variance $\Delta$ is independently
added to all the components of the message that passes through it. ${\tilde \bx}$
is first converted into ${\bx}$ which is made of $L$ sections,
each of size $B$: If the $i^{th}$ component of the original message
$\tilde \bx$ is the $k^{th}$ symbol of the alphabet, the $i^{th}$
section of $\bx$ contains only zeros, except at the position $k$, where
there is a positive value. Here, as in \cite{barbier2014replica}, we study
the simplest setting, where the non-zero value in each section
$l\in\{1,...,L\}$ is equal to $1$. As an example, if ${\tilde \bx} =
[a,c,b,a]$ where the alphabet has only three symbols $\{a,b,c\}$ then ${\bx}$ made of $4$ sections (one per symbol in the message ${\tilde \bx}$) is
${\bx}=[[100],[001],[010],[100]]$. $\bx$ is then encoded through a linear
transform by application of the operator $\bF$ of size $M\times N$ and
the resulting vector $\tilde \by$, the codeword, is sent through the Gaussian noisy
channel which outputs a corrupted version $\by$ to the receiver. The
channel thus corresponds exactly to the model given by
(\ref{eq_tildeY}), (\ref{eq_Y}) with the same notations, and is
represented in Fig.~\ref{fig_AWGN}. The rate $R$ denotes the number of
informative bits sent per channel use. Defining $K:=\log_2(B^L)$ as the number 
of informative bits in the signal $\bx$ made of $L$ sections of size $B$,
\begin{equation}
 R:=\frac{K}{M}=\frac{L \log_2(B)}{\alpha N} =\frac{\log_2(B)}{\alpha B},
\end{equation}
and therefore
\begin{equation}
M/N =\alpha=\log_2(B)/(RB) \to M=L\log_2(B)/R.
\label{eq_alphaR}
\end{equation}
The capacity $C:= 1/2\log_2(1+{\rm{snr}})$ denotes the maximum
rate achievable on the AWGN channel with power constrained codeword, where the signal to noise ratio is
${\rm{snr}}:= 1/\Delta$ (given the use of a proper rescaling of
the matrix enforcing the power constraint $||\tilde \by||_2^2=1$). The capacity is independent of the coding and decoding scheme. We also define the optimal threshold as the best achievable rate using the present superposition coding strategy with a given section size $B$ over the power constrained AWGN channel. This threshold actually tends to the capacity as $B$ increases \cite{barbier2014replica}, see Fig.~\ref{fig_phaseDiagSC}. Finally, the BP threshold is the rate until which the AMP decoder performs well combined with sparse superposition coding without the need of spatial coupling (such that it is optimal in the sense that the optimal threshold is the unique fixed point of the message passing equations). In the previous part on compressed sensing, the BP threshold was already present but as we were focusing on noiseless problems, the equivalent of the optimal threshold was identical to the MAP threshold which is the equivalent of the capacity here. In this framework, the error estimate we are interested in is the Section Error Rate (SER) which is the fraction of wrongly decoded sections: 
\begin{equation}
\label{def:SER1}
{\rm{SER}} := \frac{1}{L} \sum_{l=1}^L \mathbb{I} \left(\hat{\tilde{ x}}_l \ne \tilde{ x}_l \right),
\end{equation}
where $\hat{\tilde{ x}}_l$ is the final estimate by the AMP algorithm of the $l^{th}$ section, $l\in\{1,...,L\}$.

\subsection{AMP decoding}
Superposition  codes are interesting in the present framework because the
decoding task is actually equivalent to a multi-dimensional
sparse-estimation problem \cite{barbier2014replica}. As shown in
Fig.~\ref{fig_1dOp}, by considering the $B$-dimensional sections as the new variables, the linear
system becomes the one considered in the
compressed sensing setting. The prior $P(\cv{\bx})$ however is not (\ref{Px}),
but instead imposes that each
$B$-dimensional vector should have exactly one component equal to one and all the others equal to zero.
In other words, the vector should point exactly toward one of
the corners of the hyper-cube in $B$ dimensions.

The AMP decoder for superposition codes \cite{barbier2014replica} is
thus equivalent to the one described in Fig.~\ref{algo_AMP} with the
only difference that the thresholding functions
$f_{a_i}$ and $f_{c_i}$ are now functions of the 
sets $(\bxigma_l^{t+1},\bR_l^{t+1}):=
\{(\Sigma_i^{t+1},R_i^{t+1}):i\in l\}$. Again, $i\in l$ means that
 $x_i$ is a component of the $l^{th}$ section of
$\bx$. These functions \cite{barbier2014replica} are given by:
\begin{eqnarray}
a_i^t &:= f_{a_i}(\bxigma_l^{t+1}, \bR_l^{t+1}) = \frac{e^{-\frac{1-2R_i^t}{2(\Sigma_i^t)^2}}}{\sum_{\{j \in l\}}^B e^{{-\frac{1-2R_j^t}{2(\Sigma_j^t)^2}}}} ,\\
v_i^t &:= f_{c_i}(\bxigma_l^{t+1}, \bR_l^{t+1}) =  a_i^t (1- a_i^t ) .
\end{eqnarray}

The constraint imposed by $P(\cv{\bx})$ is much stronger than the one in compressed sensing, as it enforces binary values on the signal components and couples $B$ of them.
For this reason, perfect reconstruction might be possible even in a noisy setting.
As for compressed sensing, the asymptotic performance of
AMP for reconstruction in superposition codes is amenable to an
analytic formula \cite{barbier2014replica} using the
state evolution technique. As for real and complex variables, state evolution for
multi-dimensional variables can be demonstrated rigorously
\cite{javanmard2013state}. The analysis of AMP for i.i.d random
matrices in the context of superposition codes has been performed using the state evolution technique in
\cite{barbier2014replica}. We have repeated this analysis for
spatially coupled matrices and observed that, perhaps not
surprisingly, one could asymptotically reach capacity (see Fig.~\ref{fig_phaseDiagSC}). The object of
the next section is thus to analyze the behavior of AMP when the
i.i.d random matrices are replaced by full or spatially coupled structured operators as we did for compressed sensing.

\begin{figure}[!t]
\centering
\includegraphics[width=0.7\textwidth]{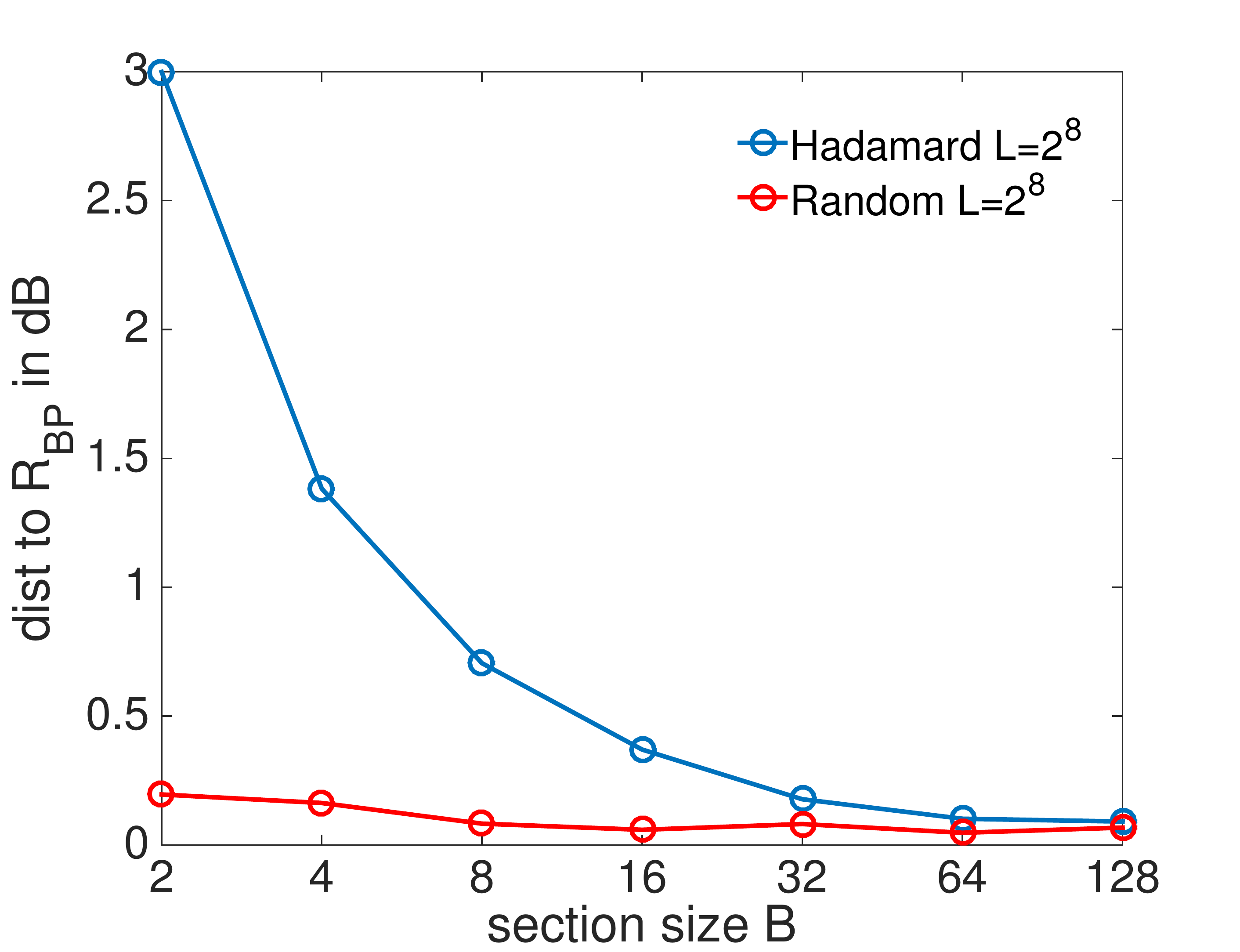}
\caption{Comparison between the distance in dB to the asymptotic BP threshold $R_{BP}$ at which the AMP decoder with full Hadamard structured coding operators (blue line) or with random i.i.d Gaussian matrices (red line) starts to reach an SER $<10^{-5}$ (which is then almost always strictly 0) for a fixed number of sections $L=2^8$ and ${\rm{snr}}=100$. The points have been obtained by averaging over 100 random instances. The BP threshold is obtained by state evolution analysis. The Hadamard operator works poorly when the signal density increases (i.e. when $B$ decreases), but gets quickly closer to the random matrix performances as it decreases. The random Gaussian i.i.d matrices have a performance that is close to constant as a function of $B$, as it should at fixed $L$.}\label{fig_distToRbp}

\end{figure}
%essentialy means to perfectly decode, as the SER is actually 0 in most of the cases if not all)
%
\begin{figure*}
\begin{centering}
\includegraphics[width=0.45\textwidth]{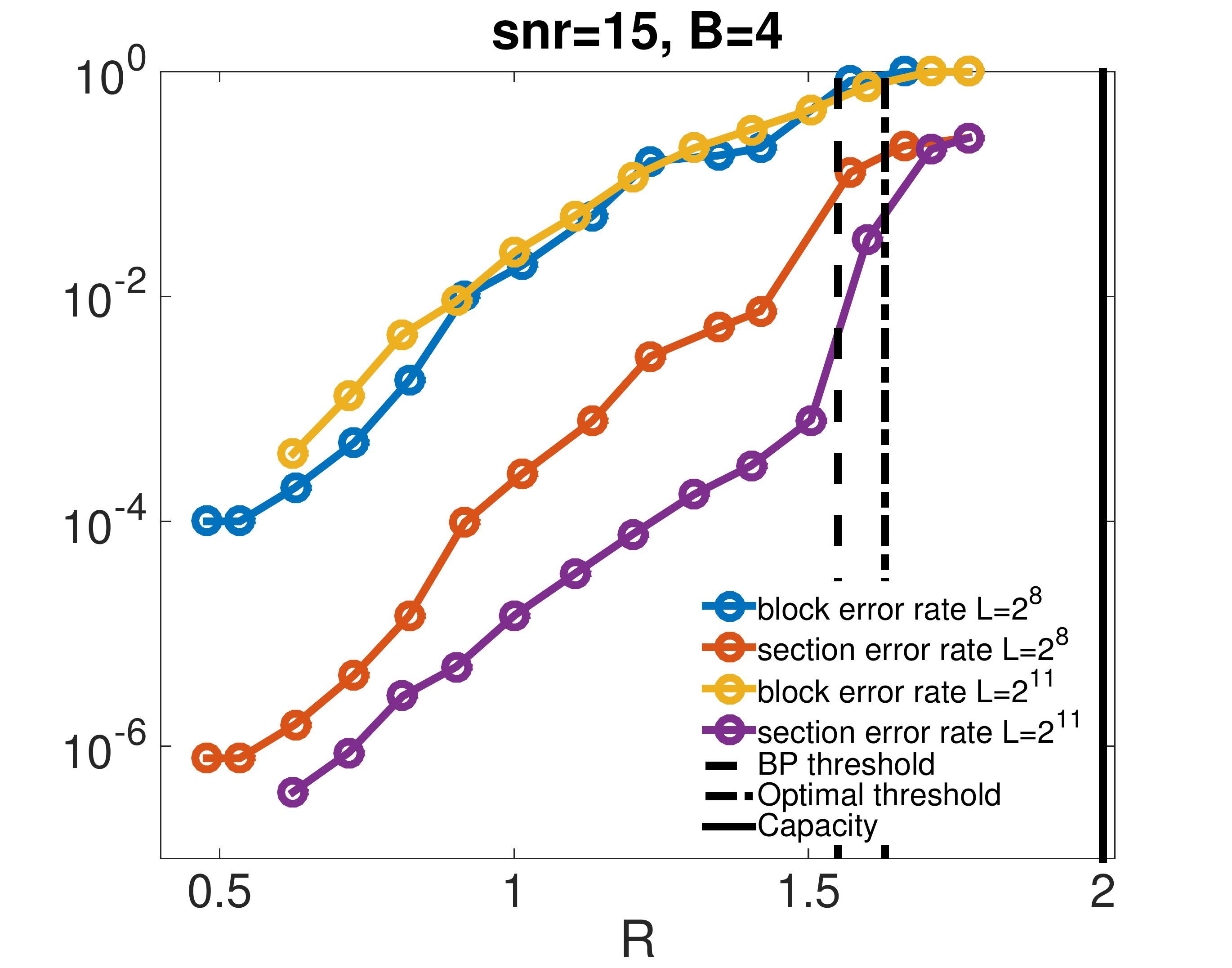}
\includegraphics[width=0.45\textwidth]{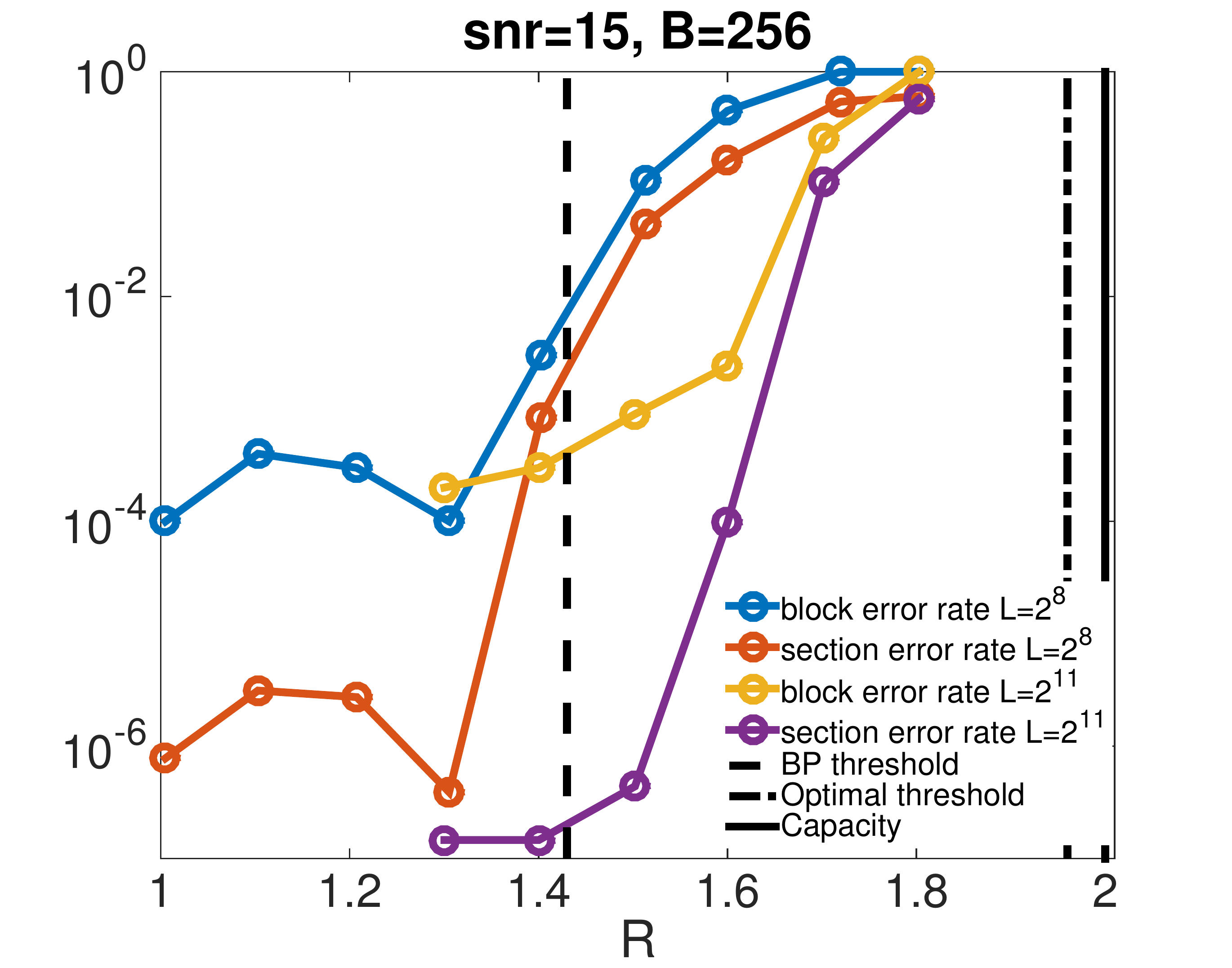}
\includegraphics[width=0.45\textwidth]{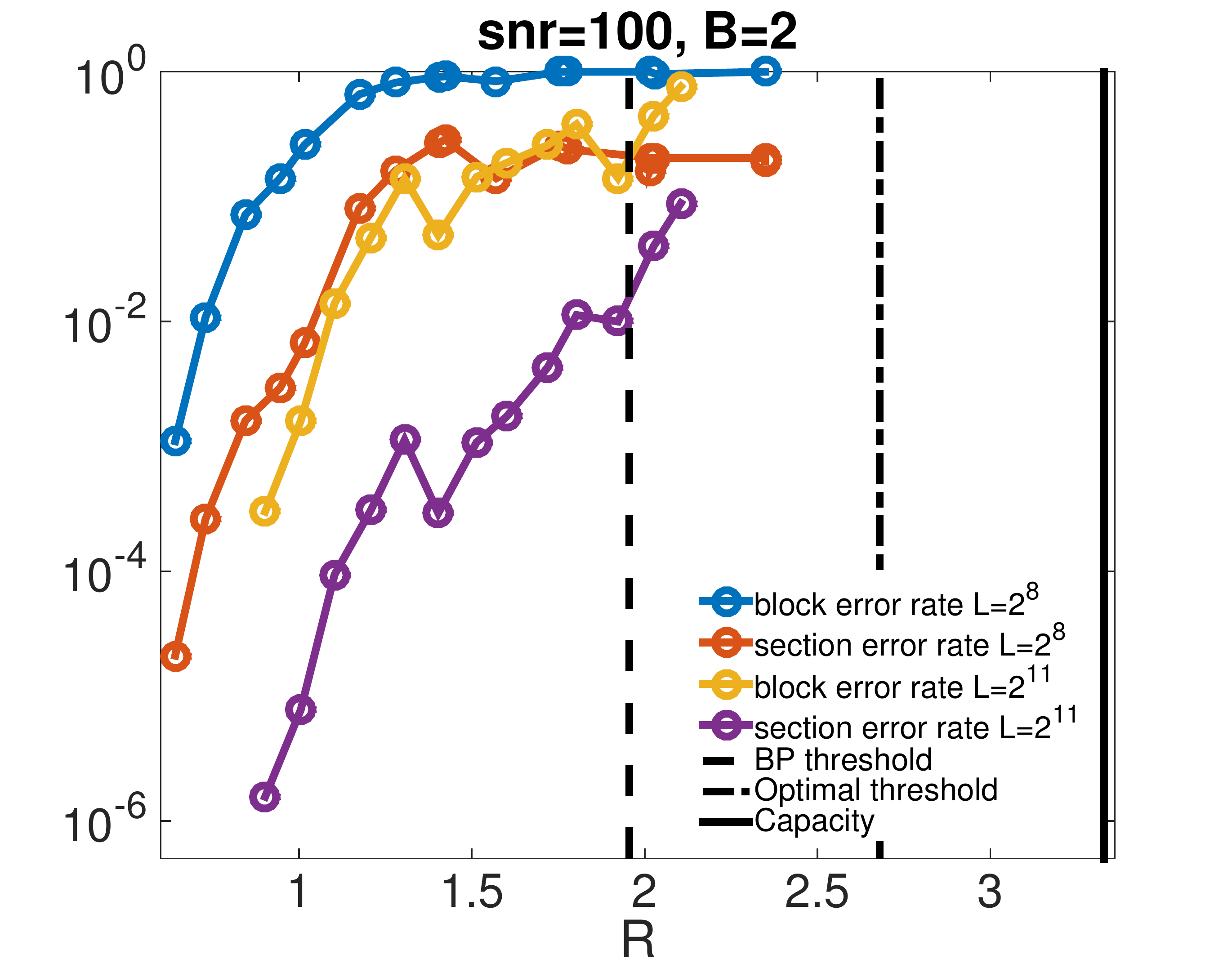}
\includegraphics[width=0.45\textwidth]{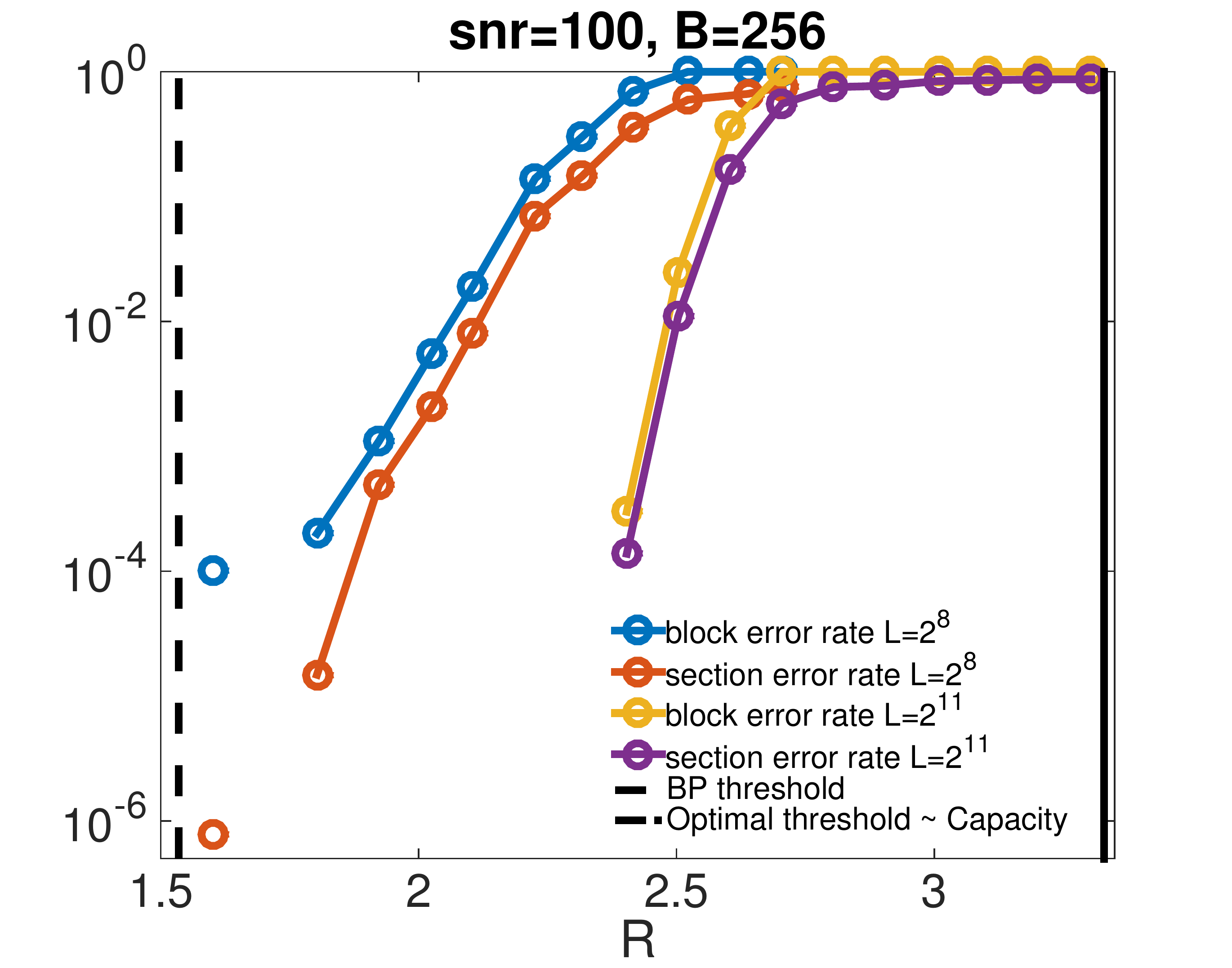}
\caption{On this plot, we show the block error and section error rates of the superposition codes using AMP combined with the spatially coupled Hadamard operator for two different ${\rm{snr}}$, two signal sizes $L$ and two section sizes $B$.
The block error rate is the fraction of the $10^4$ random instances we ran for each point that have not been perfectly reconstructed, i.e.  in these instances at least one section has not been well recontructed (the final SER $> 0$). 
The section error rate is the fraction of wrongly decoded sections (\ref{def:SER1}).
The convergence criterion is that the mean change in the variables estimates between two consecutive iterations $\delta_{\rm{max}} < 10^{-8}$ and the maximum number of iterations is $t_{\rm max}=3000$.
The upper plots are for ${\rm{snr}}=15$, the lower for ${\rm{snr}}=100$ (notice the different x axes). 
The first dashed black line is the BP threshold obtained by state evolution analysis and marks the limit of efficiency of AMP {\it{without}} spatial coupling for large signals, the second one is the Bayes optimal threshold obtained by the replica method which is the best performance any decoder can reach with superposition codes for a given section size B (obtained by the replica method, the details can be found in \cite{barbier2014replica}) and the solid black line is the capacity which bounds the performance of any coding scheme for this ${\rm{snr}}$. 
In the (${\rm{snr}}=100, B=256$) case, the optimal threshold is so close to the capacity that we plot a single line.
For such sizes, the block error rate is $0$ for
rates lower than the lowest represented rate. For $B=256$, the sharp phase
transition between the phases where reconstruction is
possible/impossible by AMP with spatial coupling is clear. The spatially coupled
operators used for the experiment are taken with parameters
$(L_c=16,L_r=L_c+1,w=2,\sqrt{J}=0.4,\beta_{{\rm{seed}}}=1.8)$, optimized heuristically.}\label{fig_finiteSizeSeeded}
\end{centering}
\end{figure*}
\subsection{Experimental results}
A first observation is that performances very quickly converge to the ones
with i.i.d random matrices with increasing alphabet size $B$. As in
CS (see section \ref{sec:results}), performances are thus asymptotically comparable. 
 Fig.~\ref{fig_distToRbp} shows the distance in dB from the BP threshold where the AMP decoder starts to get good reconstruction with i.i.d or structured matrices, as a function of $B$. It shows that structured operators reach quickly the i.i.d performances a $B$ increases.

We now turn to the behavior of spatially coupled Hadamard operators by looking at the block error rate for different settings. The block error rate is the empirical probability of not perfectly reconstructing the signal (i.e. such that the final SER $>0$). Fig.~\ref{fig_finiteSizeSeeded} shows how the finite size decreases
the performance of the spatially coupled AMP reconstruction. For the purpose of
numerical analysis, we used ($L_c=16,L_r=L_c+1,w=2,\sqrt{J}=0.4, \beta_{{\rm{seed}}}=1.8)$ and considered two different values ${\rm{snr}}=15$ and ${\rm{snr}}=100$. For $B=256$, the gain upon the standard AMP
approach with full operators is consequent, even at these finite sizes. In all the experimental settings except $({\rm{snr}}=15,B=4)$, the block error rate decreases as $L$ increases, but in the high noise one with small section size (${\rm{snr}}=15, B=4$) the error is dominated by the noise influence and the finite size sensitivity is negligible despite that the ${\rm SER}$ effectively decreases as $L$ increases. 
We observe that in the large section size $B=256$ cases, the decoder performs well at rates larger than the BP threshold as expected. When $B$ is small, the gap between the BP and optimal thresholds is too small to allow real improvement, but this gap gets larger as $B$ and the ${\rm{snr}}$ increase. In addition, we see that the gap between the optimal threshold and the capacity decreases as $B$ increases: superposition codes combined with spatially coupled operators are asymptotically capacity achieving \cite{barbier2014replica} as $B\to\infty$.
At large enough distance from $C$, the
final ${\rm{SER}}$ is exactly zero in most of the cases, giving a low block error rate. This is due to the fact that in order to observe an ${\rm{SER}} = O(\epsilon)$, there must be at
least $L= O(1/\epsilon)$ sections, which is not the case for small
signals, when the asymptotic ${\rm{SER}}$ is small.
The optimal threshold is obtained by the replica method~\cite{barbier2014replica}.

\begin{figure}[h!]
\centering
  \includegraphics[width=0.75\textwidth]{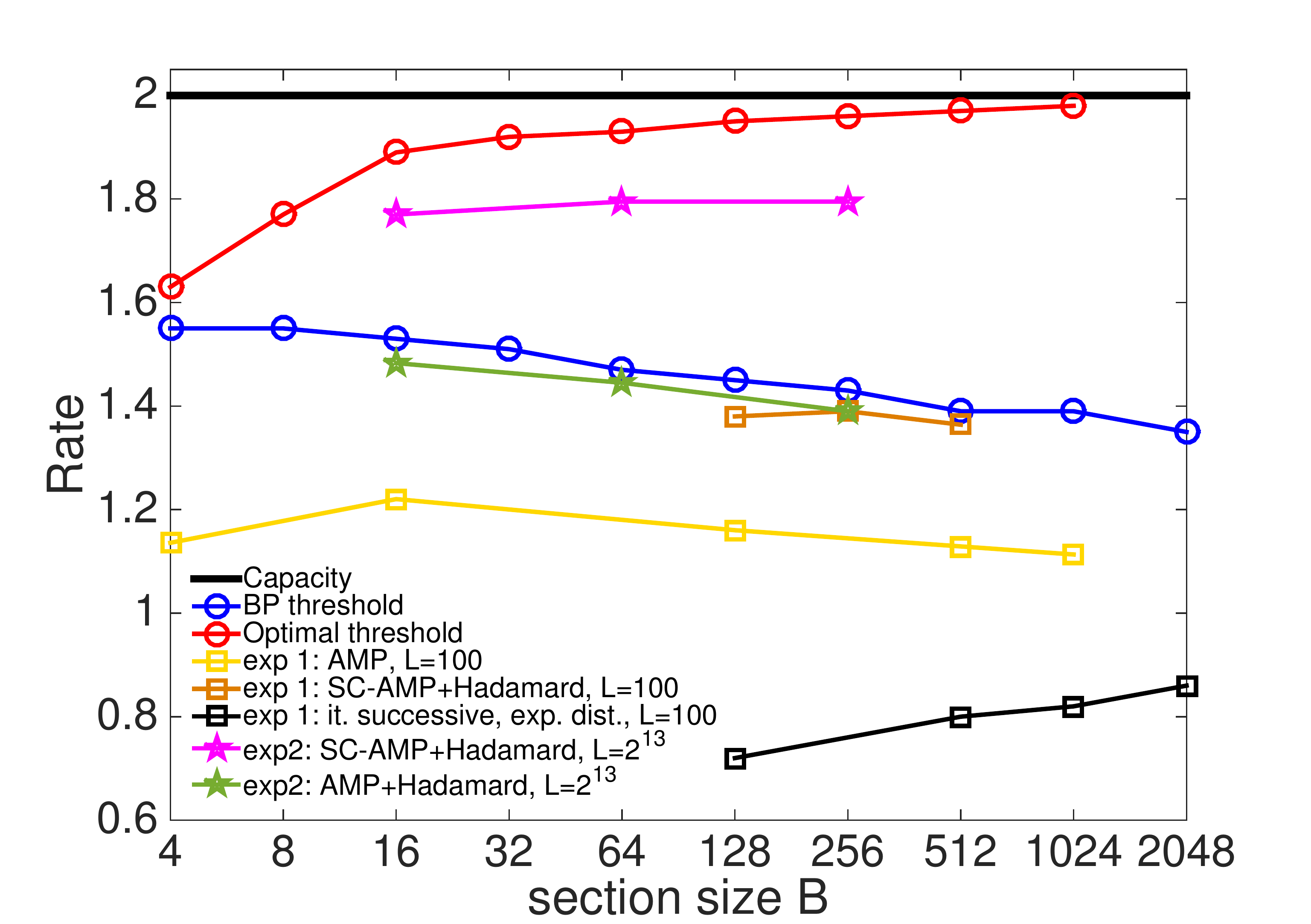}
\caption{Phase diagram and experimental results for superposition codes: Numerical experiment at finite size $L$ for
${\rm{snr}}=15$ and asymptotic results. The solid black line is the capacity, the blue line is the BP threshold obtained by state evolution analysis and the red line is the Bayesian optimal threshold (obtained by the replica method, the details can be found in \cite{barbier2014replica}).
The yellow, black and brown curves are results of the following experiment (exp 1): decode $10^4$ random instances and identify the transition line between a
phase where the probability $p_\epsilon$ to have a
${\rm{SER}}>10^{-1}$ is $p_\epsilon<10^{-3}$ (below the line) from a
phase where $p_\epsilon\ge10^{-3}$ (more than 9 instances have
failed over the $10^4$ ones). The  green and pink curves are the result of the
second protocol (exp 2) which is a relaxed version of exp 1 with $10^2$ random instances and $p_\epsilon<10^{-1}$ (below the line), $p_\epsilon\ge10^{-1}$ above.
Note that in our experiments ${\rm{SER}}<10^{-1}$
essentially means ${\rm{SER}}=0$ at these sizes. The yellow curve compares our results with the iterative successive decoder (black curve) of
\cite{barron2010sparse,barron2011analysis} where the
number of sections $L=100$. Note that these data, taken from
\cite{barron2010sparse,barron2011analysis}, have been generated with an exponential signal distribution rather than the
$\{0,1\}$ we used here. Compared with the yellow curve (AMP with the same
value of $L$) the better quality of AMP reconstruction is
clear. The green and pink curves are here to show the efficiency of the Hadamard operator with AMP with (pink curve) or without (green curve) spatial coupling.
Comparing the brown and the pink curve shows the diminishing influence of finite size effects as $L$ increases. 
The remaining gap between the pink curve and the optimal threshold is due to the still finite size of the signal.
For the experimental results, the maximum number of iterations of the algorithm is arbitrarily fixed to $t_{max}=500$. 
The parameters used for the spatially coupled operators are $(L_r=16, L_c=L_r+1, w=2, \sqrt{J}=0.3, \beta_s=1.2)$.
Tuning these parameters could further improve performances.}
\label{fig_phaseDiagSC}

\end{figure}

Fig.~\ref{fig_phaseDiagSC} shows the phase diagram
for superposition codes at fixed ${\rm{snr}}=15$ for the same
experiments as in \cite{barron2010sparse}. The rate that can be reached
is shown as a function of $B$. Comparing the black and
yellow curves, it is clear that even without spatial coupling, AMP outperforms the
iterative successive decoder of \cite{barron2010sparse} for practical
$B$'s. With the Hadamard spatially coupled AMP algorithm, this is true for any $B$ and is even more
pronounced (brown curve). The green (pink) curve shows that the full (spatially coupled) Hadamard operator has very good performances for reasonably large signals, corresponding here to a
blocklength $M<64000$ (the blocklength is the size of the transmitted vector $\tilde \by$).
\section{Conclusion}
We have presented a large empirical study using structured Fourier and
Hadamard operators in sparse estimation problems. We have shown that combining these operators with 
a spatial coupling strategy allows to reach
information-theoretical limits. We have tested our algorithm for noiseless compressed
sensing, both for real and imaginary variables, and for the decoding of sparse
superposition codes over the AWGN channel. With respect to
\cite{KrzakalaPRX2012,KrzakalaMezard12}, the resulting algorithm is
more efficient in terms of memory and running time. This allows us to deal with signal sizes as
high as $10^{6}$ and $\alpha \approx \rho$ on a personal laptop using MATLAB, and
achieve perfect reconstruction in about a minute. We have released a Matlab implementation of
our decoder at 
\mbox{\url{github.com/jeanbarbier/BPCS_common}}.
\section{Acknowledgements}
The research leading to these results has received funding from the
European Research Council under the European Union's $7^{th}$
Framework Programme (FP/2007-2013)/ERC Grant Agreement 307087-SPARCS
and from the French Ministry of defense/DGA.
\newpage
\bibliographystyle{IEEEbib}
\bibliography{refs2}

\end{document}